\documentclass[namedreferences]{SolarPhysics}

\usepackage[optionalrh,solaenum]{spr-sola-addons} % For Solar Physics
\usepackage{graphicx}                    % For eps figures, newer & more powerfull
\usepackage{amssymb}                    % useful mathematical symbols
\usepackage{color}                       % For color text: \color command
\usepackage{url}                         % For breaking URLs easily trough lines
\usepackage{twoopt}
\usepackage[breaklinks=true]{hyperref} %RR to avoid \citeads line fills
\def\rmit#1{#1}               %RR A&A & ApJ: latin abbreviations in Roman
\def\specchar#1{{\sc #1}}     %RR bad A&A style for ionization stages

\newcount\longrefs
\def\aap{\ifnum\longrefs=1 {Astron.\ Astrophys.}\else
                           {A\hbox{\rm \&}A}\fi}
\def\aapr{\ifnum\longrefs=1 {Astron.\ Astrophys.\ Rev.}\else
                            {A\hbox{\rm \&}AR}\fi}
\def\aaps{\ifnum\longrefs=1 {Astron.\ Astrophys.\ Suppl.}\else
                            {A\hbox{\rm \&}A Suppl.}\fi}
\def\aj{\ifnum\longrefs=1 {Astron.\ J.}\else
                          {AJ}\fi}
\def\ao{\ifnum\longrefs=1 {Applied Optics}\else
                           {Appl.\ Opt.}\fi}
\def\aspcs{\ifnum\longrefs=1 {Astron.\ Soc.\ Pacific Conf. Series}\else
                           {ASP Conf.\ Ser.}\fi}
\def\apj{\ifnum\longrefs=1 {Astrophys.\ J.}\else
                           {ApJ}\fi}
\def\apjl{\ifnum\longrefs=1 {Astrophys.\ J. Lett.}\else
                            {ApJ}\fi}
\def\aplett{\ifnum\longrefs=1 {Astrophys.\ J. Lett.}\else
                            {ApJ}\fi}
\def\apjs{\ifnum\longrefs=1 {Astrophys.\ J. Suppl.}\else
                            {ApJS}\fi}
\def\apss{\ifnum\longrefs=1 {Astrophys.\ and Space Science}\else
                            {Astrophys.\ Space Sci.}\fi}
\def\araa{\ifnum\longrefs=1 {Ann.\ Rev.\ Astron.\ Astrophys.}\else
                            {ARA\hbox{\rm \&}A}\fi}
\def\azh{\ifnum\longrefs=1 {Astronomicheskii Zhurnal}\else
                            {Astron.\ Zhur.}\fi}
\def\baas{\ifnum\longrefs=1 {Bull.\ Am.\ Astron.\ Soc.}\else
                            {BAAS}\fi}
\def\bain{\ifnum\longrefs=1 {Bull.\ Astronom.\ Institutes Netherlands}\else
                            {Bull.\ Astr.\ Inst.\ Neth.}\fi}
\def\gca{\ifnum\longrefs=1 {Geochim.\ Cosmochim.\ Acta}\else
                           {Geochim.\ Cosmochim.\ Acta}\fi}
\def\grl{\ifnum\longrefs=1 {Geophys.\ Res.\ Lett.}\else
                           {Geoph.\ Res.\ Lett.}\fi}
\def\iaucirc{\ifnum\longrefs=1 {IAU Circulars}\else
                          {IAU Circ.}\fi}
\def\ip{\ifnum\longrefs=1 {in press}\else
                          {in press}\fi}
\def\jgr{\ifnum\longrefs=1 {J.\ Geophys.\ Res.}\else
                           {J.\ Geophys.\ Res.}\fi}
\def\jrasc{\ifnum\longrefs=1 {J.\ Royal Astron.\ Soc.\ Canada}\else
                           {JRAS Can.}\fi}
\def\memsai{\ifnum\longrefs=1 {Mem.~Soc.~Astron.~Italiana}\else
                              {MemSAI}\fi}
\def\mnras{\ifnum\longrefs=1 {Mon.\ Not.\ Roy.\ Astron.\ Soc.}\else
                             {MNRAS}\fi}
\def\nat{\ifnum\longrefs=1 {Nature}\else
                           {Nat}\fi}
\def\pasj{\ifnum\longrefs=1 {Pub.\ Astron.\ Soc.\ Japan}\else
                            {PASJ}\fi}
\def\pasp{\ifnum\longrefs=1 {Pub.\ Astron.\ Soc.\ Pacific}\else
                            {PASP}\fi}
\def\physscr{\ifnum\longrefs=1 {Physica Scripta}\else
                            {Phys.\ Scrip.}\fi}
\def\planss{\ifnum\longrefs=1 {Planetary \& Space Science}\else
                            {Plan. \& Space Sci.}\fi}
\def\procspie{\ifnum\longrefs=1 {Proc.\ SPIE}\else
                            {Proc.\ SPIE}\fi}
\def\qjras{\ifnum\longrefs=1 {Quarterly J.\ Royal Astron.\ Soc.}\else
                            {QJRAS}\fi}
\def\sa{\ifnum\longrefs=1 {Soviet Astron..}\else
                               {Sov.\ Astron.}\fi}
\def\skytel{\ifnum\longrefs=1 {Sky \& Telescope}\else
                            {Sky \& Tel.}\fi}
\def\solphys{\ifnum\longrefs=1 {Solar Phys.}\else
                               {Sol.\ Phys.}\fi}
\def\ssr{\ifnum\longrefs=1 {Space Science Rev.}\else
                               {Space\ Sci.\ Rev.}\fi}
\def\zap{\ifnum\longrefs=1 {Zeitschr.\ f.\ Astrophysik}\else
                               {Z.\ Astrophys.}\fi}
\def\rmit#1{{\it #1}}              %% italics (RR style, Kluwer)
\def\etal{\rmit{et al.}}           %% use \etal\ for space behind it

\def\ie{\rmit{i.e.,}}              %% , required (Webster 1681)
\def\eg{\rmit{e.g.,}}              %% , required (Webster 1681)
\def\CaIIK{\mbox{Ca\,\specchar{ii}\,\,K}}       %% use \CaIIK\ for space
\def\CaIIH{\mbox{Ca\,\specchar{ii}\,\,H}}
\def\CaII{\mbox{Ca\,\specchar{ii}}}
\def\NaD{\mbox{Na D$_1$}}

\def\Hmin{\hbox{\rmH$^{^{_-}}\!$}}      %% H^min, very elegant
\def\HK{\mbox{H\,\&\,K}}
\def\HtwoV{\mbox{H$_{2V}$}}
\def\KtwoV{\mbox{K$_{2V}$}}
\def\kms{\hbox{km$\;$s$^{-1}$}}
\def\gf{\mbox{$g \! f$}}
\def\is{\!=\!}                             %% tighter spacing
 \def\rmH{{\rm H}}

\begin{document}
\begin{article}
\begin{opening}

\title{The wings of Ca\,II H \& K as photospheric diagnostics and
the reliability of one-dimensional photosphere modeling}

%%%%%%%%%%%%%%%%%%%%%%%%%%%%%%%%%%%%%%%%%%%%%%%%%%%
%% Authors Names
%
\author{V.A. \surname{Sheminova}           }

%%%%%%%%%%%%%%%%%%%%%%%%%%%%%%%%%%%%%%%%%%%%%%%%%%%
%% Runningheads
%
\runningauthor{V.A. Sheminova}
  \runningtitle{Ca\,II H\&K wing diagnostics}

%%%%%%%%%%%%%%%%%%%%%%%%%%%%%%%%%%%%%%%%%%%%%%%%%%%
%% Affilations
%
  \institute{Main Astronomical Observatory, National Academy of
  Sciences of Ukraine, 27 Akademika Zabolotnoho St., Kiev, 03680
  Ukraine\\
                       email: \url{shem@mao.kiev.ua}
  }

%%%%%%%%%%%%%%%%%%%%%%%%%%%%%%%%%%%%%%%%%%%%%%%%%%%
%%% Abstract
\begin{abstract}

  The extended wings of the \CaII\ \HK\ lines provide excellent
    diagnostics of the temperature stratification of the photosphere
    of the Sun and of other cool stars, thanks to their LTE opacities
    and source functions and  their large span in formation height.
    The aim of this study is to calibrate the usage of the \HK\ wings
    in one-dimensional  interpretation of spatially averaged spectra
    and in deriving per-pixel stratifications from resolved spectra.
  I use multi-dimensional simulations of solar convection to
    synthesize the \HK\ wings, derive one-dimensional models from
    these wings as if they were observed, and compare the resulting
    models to the actual simulation input.
  I find that spatially-averaged models constructed from the
    synthesized wings generally match the simulation averages well,
    except for the deepest layers of the photosphere where large
    thermal inhomogeneities and Planck-function nonlinearity gives large errors.
    The larger the inhomogeneity, the larger the latter.
    The presence of strong
    network fields increases such inhomogeneity.  For quiet  photospheric
    conditions the temperature excesses reach about 200~K.  One-dimensional
    stratification fits of discrete structures such as granulation and
    small-scale magnetic concentrations give satisfactory results with
    errors that are primarily due to steep temperature gradients and
    abrupt changes of temperature with depth.
  I conclude that stratification modeling using the \HK\ wings is
    a useful technique for the interpretation of solar high-resolution
    observations.

\end{abstract}

%%%%%%%%%%%%%%%%%%%%%%%%%%%%%%%%%%%%%%%%%%%%%%%%%%%
%% Keywords
%

\end{opening}
%-------------------------------------------------

%%%%%%%%%%%%%%%%%%%%%%%%%%%%%%%%%%%%%%%%%%%%%%%%%%%
%% Sections

%%%%%%%%%%%%%%%%%%%%%%%%%%%%%%%%%%%%%%%%%%%%%%%%%%%%%%%%%%%%%%%%%%%%%%%%%%%%
\section{Introduction}                              \label{sec:introduction}
%%%%%%%%%%%%%%%%%%%%%%%%%%%%%%%%%%%%%%%%%%%%%%%%%%%%%%%%%%%%%%%%%%%%%%%%%%%%
The wings of the \CaII\ \HK\ lines at 3933\,\AA\ and 3968\,\AA\
span the solar photosphere in their formation which is
relatively simple to interpret.  They contain suitable
blend-free windows to sample intensities as well as blends to sample
Dopplershifts at different heights. The span of their height
sampling is extraordinary wide.
\inlinecite{1989SoPh..124...15A}  %? Ayres 400nm as deep as 1.6mu
demonstrated that the opacity minimum at 4000\,\AA\ and the
large short-wavelength Planck sensitivity to temperature make
the \HK\ background continuum respond as deep as the \Hmin\
opacity minimum near 1.6\,$\mu$m.  The \HK\ cores are
chromospheric at line center and sample the shock-ridden
internetwork ``clapotisphere'' as \HtwoV\ and \KtwoV\ grains
near line center
(\eg\ \opencite{1995ESASP.376a.151R}). % Rutten Asilomar clapotisphere
The extended wings sample all heights between these extremes,
with clean single-peaked contribution functions because
virtually all calcium resides in the \CaII\ ground state without
sensitivity to Boltzmann and Saha population partitioning
(Figure~6 of
\opencite{2006A&A...449.1209L}) % Leenaarts++ BPs blue wing Ha
and without sensitivity to NLTE population processes such as
photon pumping or photon suction.  The source functions of \HK\
also obey LTE--CRD until about 1\,\AA\ from line center (\eg\
\opencite{1975ApJ...199..724S}; % Shine+Milkey+Mihalas, CaK PRD
\opencite{1975ApJ...201..799A}; %T Ayres CS wing modeling
\opencite{1989A&A...213..360U}). %T Uitenbroek, PRD CaII
\inlinecite{1980ApJ...241..448O} emphasized the closeness  of
its  source function to LTE which results from the relatively high
probability  of wing-photon destruction by incoherent scattering
into the thermal line core.

\HK\ wing modeling of solar and stellar photospheres was
initiated during the 1970s at Boulder by J.L~Linsky and
collaborators, in particular T.R.~Ayres
(\opencite{1974ApJ...192...93A}; %T Ayres++ Procyon
 \opencite{1975ApJ...200..660A}; %T Ayres+Linsky Arcturus
 \opencite{1975ApJ...201..799A}, %T Ayres CS wing modeling
 \citeyear{1977ApJ...213..296A}). %T Ayres reexamination wing models
Later,
\inlinecite{2002A&A...389.1020R} % Rouppe penumbra wing HK
used the technique on a sunspot penumbra.  The original
application to solar faculae by
\inlinecite{1974SoPh...37..145S} %T Shine+Linsky, K wing faculae model
was followed by \citeauthor{2005A&A...437.1069S} % T Sheminova++
(\citeyear{2005A&A...437.1069S}, henceforth Paper~I) to derive
models of individual magnetic concentrations (``fluxtubes'')
from high-resolution \HK\ spectra.  \CaIIH\ wing modeling using
an inversion SIR-code based on response functions was initiated
by \inlinecite{2008A&A...479..213B}. It serves to interpret
large sets of spatially resolved line profiles in an automated
manner. Recently, \inlinecite{2009MmSAI..80..639H} % Henriques wings H
briefly described an application to high-resolution  \CaII\ H
filtergrams, including calibration using snapshots from
numerical multi-dimensional simulations.

Even more recent are the new results of
\inlinecite{2011ApJ...736...69U} who demonstrate that the
analysis of average spectra in terms of one-dimensional models
is inherently unreliable because the inferred temperatures
differ from the actual averaged temperatures.  The average
continuum of an atmosphere with temperature fluctuations is
higher than the continuum of a one-dimensional atmosphere with
the same average temperature stratification, due to the
non-linearity of the Planck function with temperature.  The
interpretation of an average spectrum from a horizontally
inhomogeneous atmosphere in terms of a one-dimensional model
therefore leads to an overestimate of the average temperatures
in the atmosphere.  Since most of the information about solar
and stellar atmospheres is obtained from unresolved spectra, it
is important to assess this erroneous temperature excess in
one-dimensional  modeling of a mean spectra.

In the present paper the usage of the \HK\ wings as
model-construction diagnostics is calibrated in an effort
similar to  the analysis of \inlinecite{2009MmSAI..80..639H} and
\inlinecite{2010mcia.conf..511H}. The first goal of the paper is
to estimate the reliability of one-dimensional modeling based on
the \CaII\ \HK\ wings for characteristic fine structures of the
solar photosphere. The second goal is to quantify the
temperature excesses produced by one-dimensional \HK\ wing
modeling from unresolved spectra.

Three different numerical simulations are employed here:
\begin{itemize}

\item the 3D hydrodynamic (HD) simulation with  the CO$^5$BOLD code
described by
  \inlinecite{2004A&A...414.1121W}, % Wedemeyer++
henceforth called 3DHD;

\item a 2D magnetohydrodynamic (MHD) simulation described  by
 \inlinecite{2000KFNT...16...99G} and
 \inlinecite{2001SoPh..203....1G}, % Gadun: 2D magnetoconvection
  henceforth 2DMHD;

\item a snapshot from a 3D magnetohydrodynamic simulation with the code of
  \inlinecite{1998ApJ...499..914S}, % Stein+Nordlund, I general properties
  henceforth 3DMHD.
  This snapshot was also used by
  \inlinecite{2004ApJ...610L.137C}, % Carlsson++ G-band BP limg
  Leenaarts \etal\
 (\citeyear{2006A&A...452L..15L}, %  Leenaarts++ BP diagnostics
 \citeyear{2006A&A...449.1209L}), %  Leenaarts++ BPs blue wing Ha
  \inlinecite{2006ApJ...648..741T}, % Uitenbroek+Tritschler BPs
  \inlinecite{2007ApJ...668..586U}, %Uitenbroek++ The Discrepancy in G-Band Contrast
  and
  \inlinecite{2009MmSAI..80..639H}. % Henriques wings H
\end{itemize}
For each snapshot I computed emergent absolute intensities
throughout \HK\ and then used these as quasi-observed \HK\
spectra to derive a best-fit one-dimensional photosphere model.
I do not consider any actual \HK\ observation in this paper, but
compare these simulation-result fits to the actual
stratifications in the original simulation snapshots and compare
the different simulations as well.

In Section 2 I describe the simulation atmospheres, the spectral
synthesis, and the procedure of one-dimensional model
construction. The synthesized \HK\ spectra are presented in
Section 3.1. The best-fit temperature stratifications of the
average atmosphere inferred from spatially averaged synthetic
\HK\ profiles and the evaluation of the temperature excesses are
presented in Section 3.2. Best-fit temperature stratifications
for granules, lanes and magnetic concentrations derived from
spatially resolved \HK\ profiles are given in  Section 3.3.
Discussion and conclusions are given in Sections 4 and 5.

%%%%%%%%%%%%%%%%%%%%%%%%%%%%%%%%%%%%%%%%%%%%%%%%%%%%%%%%%%%%%%%%%%%%%%%%%%%%
\section{Method}
%%%%%%%%%%%%%%%%%%%%%%%%%%%%%%%%%%%%%%%%%%%%%%%%%%%%%%%%%%%%%%%%%%%%%%%%%%%%

\subsection{Simulations}     \label{sec:data}
%%%%%%%%%%%%%%%%%%%%%%%%%%%%%%%%%%%%%%%%%%%%%%%%%%%%%%%%%%%%%%%%%%%%%%%%%%%%

The 3DHD simulation reproduces photospheric granulation during a
few hours of solar time. The simulation cubes measure $(x,y,z)
\is 5600\times5600\times3110$~km with an atmospheric extent of
1710~km in $z$ above  the zero ``surface'' level defined by mean
optical depth $\log\tau_5=0$  at 5000\,\AA.  The grid step is
40~km in $x$ and $y$; the height sampling in $z$ starts at 46~km
at the base and is 12~km for $z>-270$~km.  I used 55 simulation
cubes at 60~s time steps to obtain averaged \HK\ profiles.

The 2DMHD simulation is representative of magnetogranulation
with strong magnetic field, similar to network regions with
fluxtube field strengths reaching 1.5--2~kG at the surface. The
horizontal extent is about 4000~km, the vertical extent 735~km
above and 1370~km below the zero ``surface'' level.  The
start-off initial magnetic field had  bipolar geometry with loop
connectivity and spatially averaged unsigned field strength
54~G, which increased with the simulation run time to a
time-average of about 400~G at the surface.  I used 52 snapshots
at 30-s intervals.

The 3DMHD simulation snapshot measures
$6000\times6000\times5000$~km with 2500~km atmosphere height
above $\log\tau_5=0$. The grid step is 25~km horizontally,
vertically 15~km in the upper layers and 35~km in the lower. The
initial seed magnetic field was vertical and uniform with
strength 250~G. The evolved snapshot contains magnetogranulation
with a few fluxtubes in intergranular lanes with field
strengths of about 1700~G.  Weak horizontal magnetic fields
occur in the granules.

\subsection{Spectral synthesis}
%%%%%%%%%%%%%%%%%%%%%%%%%%%%%%%%%%%%%%%%%%%%%%%%%%%%%%%%%%%%%%%%%%%%%%%%%%%%

I modified the SPANSAT code described by
\inlinecite{1988ITF...87P....3G} % Gadun+Sheminova SPANSAT
for LTE spectral synthesis of the overlapping \HK\ lines
including superimposed blends.  It computes intensity spectra
containing any selection of overlapping spectral lines from a
given one-dimensional model atmosphere.  The spectral line
parameters were taken from the
VALD database \cite{1999A&AS..138..119K}.  % Kupka++ VALD
The atomic parameters of the \HK\ \CaII\ lines are the same in
Paper~I, with calcium abundance $A_{\rm Ca} \is 6.38 $ and $\log
\gf \is 0.134$ for \CaIIK, $-0.18$ for \CaIIH.

Micro- and macroturbulent velocities have small effect on the
extended \HK\ wings.  I did not apply micro and macroturbulence
in the line synthesis from the simulations since their velocity
field is given with the data cubes.  In the subsequent fitting
procedure I applied microturbulence  in the line profile
synthesis  following
\inlinecite{1986SoPh..106..237G} %Gurtovenko+Sheminova 'Crossing' method
with   the total velocity amplitude $ V_t=1.6$~\kms\ determined
as $ V_t^2=V_{\rm mic}^2+V_{\rm mac}^2$.  This approach
simplifies the \HK\ profile calculation without loss of
accuracy.

Collisional damping by neutral hydrogen atoms was formerly a
major uncertainty which has been resolved by the code of
\inlinecite{1998MNRAS.300..863B} % Barklem+OMara Ca II etc damping
based on quantum-mechanical estimation.  I used it here.

In Paper~I, which compared observed and computed \HK\ wing
profiles, we had to cope with the ``line haze'' of unresolved
weak lines in this part of the spectrum, for which we applied
ad-hoc scaling factors
\begin{equation}\label{eq_1}
  f = 1 + c \,(N_{\rm H} {/} N_{\rm H^{-}})
\end{equation}
to the computed continuum opacity. The constants $c$ were
determined from comparing synthetic far wings computed for disk
center to the observed far wings.  Since the present study does
not involve actual observations this contribution is neglected
here.

In addition, no profile comparisons are made within 1\,\AA\ from
line center to avoid effects from coherent scattering.  The
effects of magnetic fields on \HK\ wing formation are neglected.
All 2912 overlapping lines of neutral and single ionized atoms
in the VALD list are entered as blends. In between these, I
selected 28 relatively blend-free wing windows as the sampling
wavelengths that define what I call ``\HK\ profile'' here.  Such
profiles were synthesized for all columns of all simulation
snapshots. Then the average line profile was computed by
averaging all these individual profiles per simulation.

The major assumption in this computation is one-dimensional LTE
radiative transfer along columns.
\inlinecite{1980ApJ...241..448O} have shown that the unresolved
Mg~$k$ and Ca~K wings are not sensitive to the horizontal
transfer effects and may be used  as potentially good
atmospheric diagnostics in the multi-component (1.5D) sense
without significant errors. More recently,
\inlinecite{2010ApJ...709.1362L} %Leenaarts etal (2010)
have demonstrated  that 3D radiative transfer effects only
become noticeable at much larger height, as for example in the
3D evaluation of \NaD\ formation.

\subsection{One-dimensional model construction}     \label{sec:methods1}
%%%%%%%%%%%%%%%%%%%%%%%%%%%%%%%%%%%%%%%%%%%%%%%%%%%%%%%%%%%%%%%%%%%%%%%%%%%%

As in the classical \HK\ wing modeling I apply trial-and-error
fitting of the synthetic line profiles by changing the
temperature stratification with height in the atmosphere as for
a plane-parallel atmosphere.

The initial start-off model gives the temperature $T(h)$ and gas
pressure $P_g(h)$ stratifications depending on the geometric
height $h$ (positive outwards).  For the mean atmosphere
modeling and for fitting granular fine structure I chose the
HSRA-SP-M model as the start-off stratification. This is the
HSRA model of
\inlinecite{1971SoPh...18..347G} % HSRA
modified by \inlinecite{1974SoPh...34..277S} % Spruit, HSRA-SP model
and again modified in Paper~I. Note that HSRA-SP-M contains very deep
layers of the atmosphere. The PLANEW reference model for a strong
magnetic fluxtube by
\inlinecite{1992A&A...262L..29S} % Solanki+Brigljevic (1992)
was chosen as the initial model for the modeling of small-scale
magnetic elements.

In each iteration the temperature stratification was changed manually
at 14--17 geometrical height samples. Their distribution with height
is irregular and was selected depending on the temperature gradients
for the simulated object. The initial set of the height points (in
kilometers) for the start-off HSRA-SP-M was the following: 864 (-6.7),
300, 200, 150, 100, 75, 50, 25, 0, -25, -50, -100, -150, -300, -409
(3.5). For PLANEW as start-off the set was 659 (-6.2), 300, 200, 150,
100, 75, 50, 25, 0, -25, -50, -100, -150, -300, -306 (0.8). The highest
and deepest optical depths are given between brackets, in logarithmic
units.  The temperature at other heights was computed by spline
interpolation.  The resulting best-fit stratifications extend down to
the deepest geometrical depth.  Since the deepest optical depth
depends on the specific opacity $\kappa_\lambda (T,~P_g)$ it depends
on the simulated feature.

At each temperature modification a new gas pressure
stratification was computed assuming vertical hydrostatic
equilibrium. The equation of hydrostatic equilibrium can be
integrated in many ways. I used:
\begin{equation}\label{eq_2}
 P_g(h)= P_0~\exp~[-\frac{g}{R}\int_{h_0}^h
 \frac{\mu_0(h')}{T(h')}\,{\rm d}h'],
\end{equation}
where $g$ is the gravitational acceleration, $R$ the gas
constant,  $\mu_0(h)$ the variable mean molecular weight taking
into account the partial ionization of the various chemical
elements. $P_0$ is defined as the gas pressure at the bottom of
the atmosphere ($h_0$) accounting for the adopted chemical
composition. The SPANSAT code calculates the electron pressure
$P_e$ as a function of depth from $P_g(h)$ and the equation of
state for ideal gas including LTE evaluation of the ionization
equilibrium for all pertinent elements according
to the scheme by \inlinecite{1967MetCP...7....1M} % Mihalas
with minor changes.  In this work the chemical composition
corresponds to data from the paper of
\inlinecite{2007ApJ...667.1243F}.  It is given as a table of
abundance values for 104 chemical elements and it does not
change during the fitting procedure. Thus, the  free atmospheric
parameters are $(14-17) \times T $, $P_0$ in  our
one-dimensional modeling.

As in other inversion methods each subsequent iteration
minimizes the sum of the squared differences between the
original and synthetic \HK\ profiles for 28 wavelength points.
The procedure of the minimization is complex due to the
nonlinear dependence of the emergent intensity on the
atmospheric parameters that control the radiative transfer.  To
succeed one must gain experience in how the intensity of the
\HK\ profile responds to temperature changes of specific layers
of the photosphere.

Such fitting was done for the spatial average of each simulation
and for single simulation pixels for which the modeling resolved
specific fine structure (granule, intergranular lane,
small-scale magnetic element).

%------------------------------------------------------------- Fig1
\begin{figure}
   \centering
   \includegraphics[width=11.5cm]{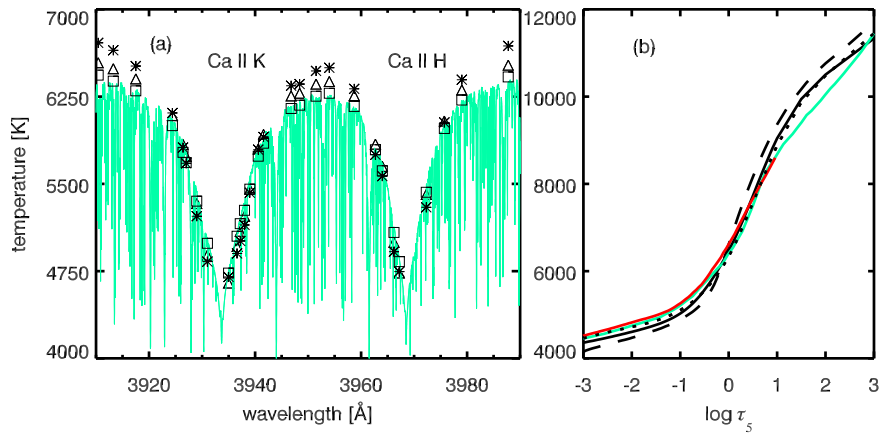}
   \caption[]{(a) Mean \HK\ profiles synthesized from 3DHD (triangles),
     2DMHD (asterisks), and 3DMHD (squares).  The green observed profile
     with numerous blends is copied from the the disk-center
     Brault-Neckel atlas by  \inlinecite{Neckel1999}. %T SP, FTS atlases.
       The intensity units were converted into brightness
       temperature. (b) Mean temperature stratification for the 3DHD
       (solid), 2DMHD (dashed), and 3DMHD (dotted) simulations.  The
       spatial averaging was done over surfaces of equal optical depth
       $\tau_5$. The temperature stratification of the quiet
       photosphere derived from unresolved observed spectra
       HSRA-SP-M (Paper 1) and  SRPM-305 (\opencite{2007ApJ...667.1243F})
        is presented by  green and red curves, respectively.  }
      \label{Fig:hk_fit}
\end{figure}
%_____________________________________________________________

%====================================================Fig2===================
\begin{figure}
   \centering
   \includegraphics[width=4.cm]{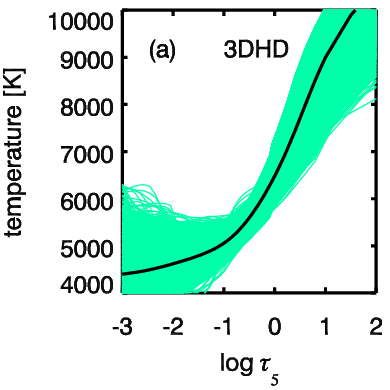}
   \includegraphics[width=3cm]{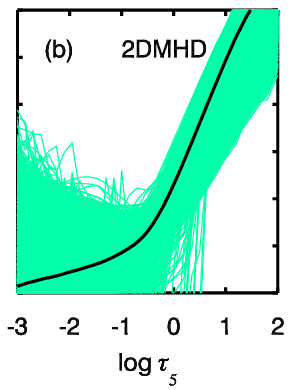}
   \includegraphics[width=3cm]{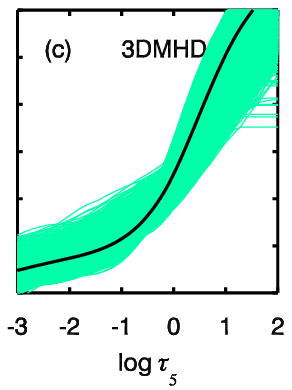}
   \caption[]{ Temperature against optical depth for all
     columns of a snapshot of the 3DHD  and 3DMHD
     simulations and for all columns of the time series of 2DMHD.
     The solid curve is the mean.
     }
   \label{Fig:t_av}
\end{figure}
%===========================================================================

%%====================================================fig3=======================
\begin{figure}
   \centering
   \includegraphics[width=6.cm]{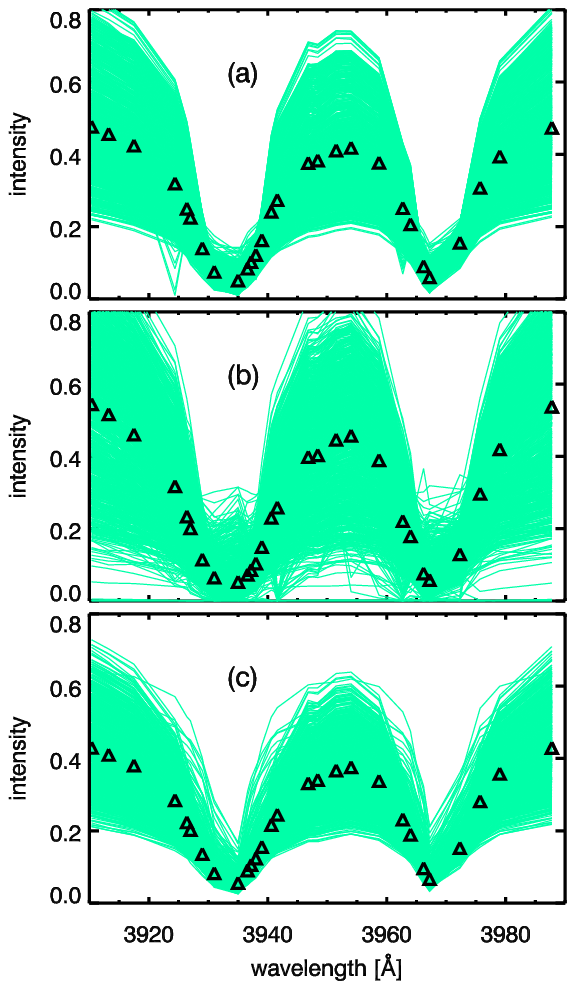}
    \includegraphics[width=5.7cm]{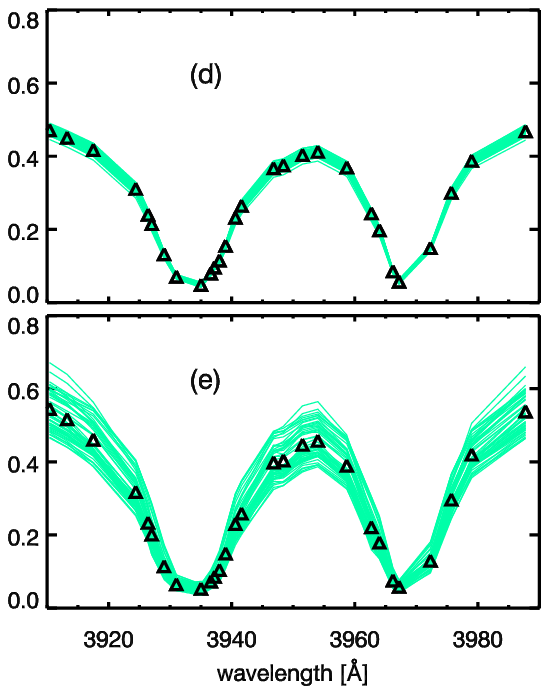}
    \caption[]{ Synthetic \HK\ profiles for all columns of all snapshots
      of 3DHD (a), 2DMHD (b), and a snapshot of 3DMHD (c).  Temporal variations
      between profiles from all snapshots of the 3DHD (d) and
      2DMHD (e) simulations.  The symbols specify the mean value.
      In this and other
      figures the absolute intensity is given in $ \rm{watt} \,
      \rm{cm}^{-2} \, \rm{ster}^{-1} \, {\AA} ^{-1}$.  }
   \label{Fig:spread}
\end{figure}
%===========================================================================

%%%%%%%%%%%%%%%%%%%%%%%%%%%%%%%%%%%%%%%%%%%%%%%%%%%%%%%%%%%%%%%%%%%%%%%%%%%%
\section{Results}                                        \label{sec:results}
%%%%%%%%%%%%%%%%%%%%%%%%%%%%%%%%%%%%%%%%%%%%%%%%%%%%%%%%%%%%%%%%%%%%%%%%%%%%

\subsection{Synthesized spectra}
%%%%%%%%%%%%%%%%%%%%%%%%%%%%%%%%%%%%%%%%%%%%%%%%%%%%%%%%%%%%%%%%%%%%%%%%%%%%

Figure~\ref{Fig:hk_fit}a shows the averaged \HK\ profiles
synthesized from all columns of the 3DHD, 2DMHD, and 3DMHD
simulations. For the first two simulations the profile averaging
is done over all columns of all snapshots, i.e., over both
surface and time. In the third case the averaging is done over
all columns of the single snapshot, i.e., only over surface. For
comparison Figure~\ref{Fig:hk_fit}a also shows observed \HK\
profiles from the disk-center atlas taken by J.W.~Brault with
the NSO Fourier Transform Spectrometer and made
available by \inlinecite{Neckel1999} %T SP, FTS atlases "announcement"
after calibration to the absolute intensities of
\inlinecite{1984SoPh...90..205N}. %T Labs+Neckel including Brault atlas
The three synthesized profiles agree reasonably with the atlas
spectrum.  However, the outer wings of the 2DMHD profile reach
appreciably larger intensities than the other profiles.

Figure~\ref{Fig:hk_fit}b shows the average temperature
stratification for the three simulations, \ie\ averaging both
spatially and temporally for 3DHD and 2DMHD, spatially only
3DMHD.  The spatial average is evaluated over surfaces with
equal continuum optical depth $\tau_5$ as in
\inlinecite{2011ApJ...736...69U}. % Uitenbroek++.
They showed that such equal-$\tau_5$ averaging represents
emergent intensities better than averaging over planes of equal
geometrical depth.

%==================================================fig4=======================
\begin{figure}
   \centering
\includegraphics[width=6.cm]{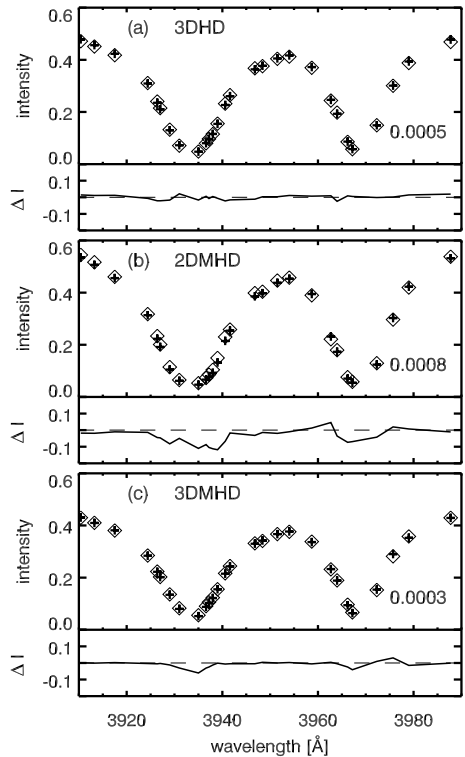}
   \caption[]{Best-fit \HK\ profiles (crosses) compared
     to the temporal and spatial averages of
     the synthetic profiles (rhombuses) from  the 3DHD (a),
     2DMHD (b), and 3DMHD (c) simulations.
     The lower panels of each
     subfigure show residuals $\Delta I$~=~$(I-I_{\rm original})/I_{\rm
     original}$, i.e., the differences between the
     synthetic and original profiles in the relative units.
     The standard deviation $\sigma(I-I_{\rm original}) $
     are indicated in the lower-right corner of each subfigure.}
         \label{Fig:best_fit}
\end{figure}
%===========================================================================

%======================================================fig5================
\begin{figure}
  \centering
     \includegraphics[width=10.cm]{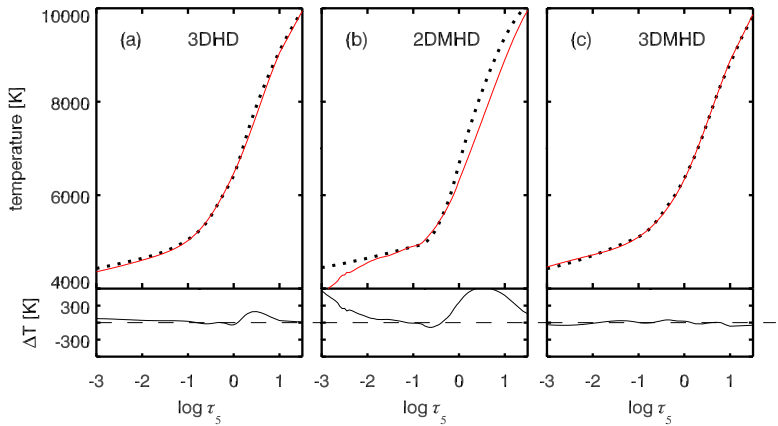}
     \caption[]{ Temperature stratifications of the best-fit models
       (dotted) compared to the equal-$\tau_5$ averages of the
       simulations (red curve). The lower panels present the residual
       temperatures $\Delta T=T_{\rm {best-fit}} - T_{\rm original}$.}
   \label{Fig:fit_mod}
\end{figure}
%===========================================================================

As may be seen from  Figure~\ref{Fig:hk_fit}b, in the deep
layers below $\log\tau_5 \is 0$ the mean gradient is smallest
for 3DMHD and largest for 2DMHD. A steeper temperature gradient
in deep layers produces larger emergent intensity in the outer
wings. For comparison Figure~\ref{Fig:hk_fit}b adds the
one-dimensional \HK\ line-fitting HSRA-SP-M model (Paper 1) used
here to start the fitting procedure and the standard
one-dimensional
continuum-fitting SRPM-305 model of \inlinecite{2007ApJ...667.1243F}.  %   SRPM 305.
Their temperature stratifications have lower gradients compared
with the simulation averages.

Figure~\ref{Fig:t_av} demonstrates the spread around the mean
temperature stratification for each simulation. The spread
caused by thermal inhomogeneities differs substantially between
the three simulations. In high layers they are largest for 2DMHD
and remarkably small for 3DMHD; in deep layers they are more
similar.

Figures~\ref{Fig:spread}a,b,c show the spatial spread in the
\HK\ intensities computed from the simulations, using all
snapshots. In each case, the spread increases towards the wings
with larger temperature spread at larger formation depth.

Figures~\ref{Fig:spread}d,e show the temporal variations between
snapshots for 3DHD and 2DMHD.  The variability is largest for
2DMHD, probably reflecting growth in inhomogeneity because the
magnetic field strength and oscillations increase with time.

%===================================FIG6====================================
\begin{figure}
   \centering
   \includegraphics[width=6.5cm]{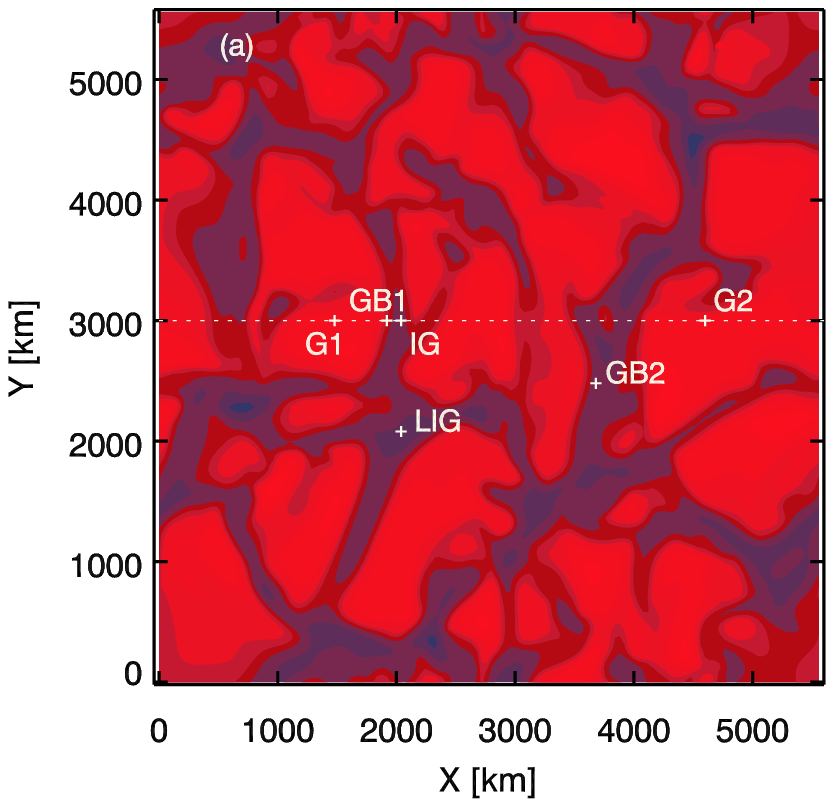}
    \includegraphics[width=5.cm]{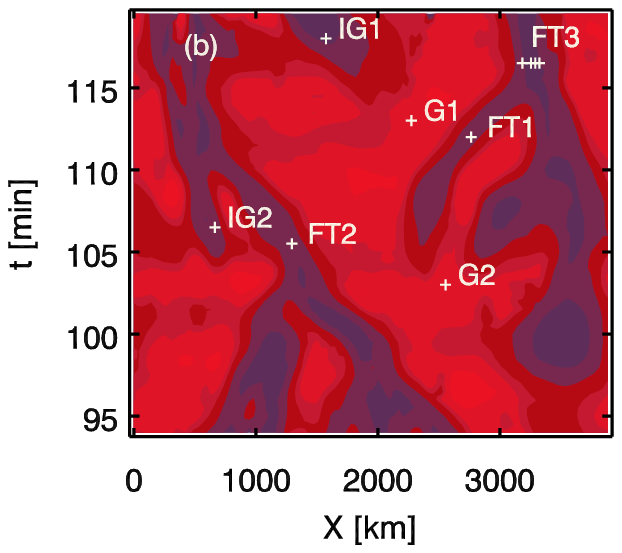}
    \includegraphics[width=7.cm]{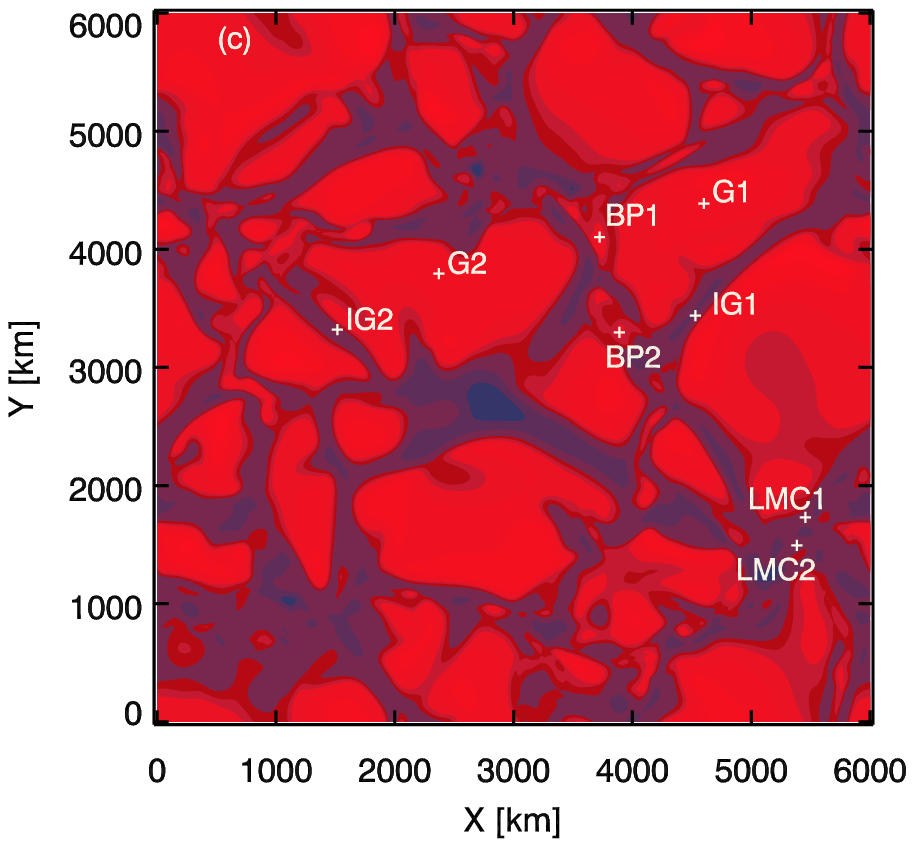}
    \caption[]{Continuous intensity images at
      $\lambda=3910$~\AA\ for a 3DHD snapshot (a), the time series of
      2DMHD (b), and a 3DMHD snapshot (c).  The locations
      selected for feature modeling
      are marked with crosses and identifiers  with designations: G1
      and G2 for granule centers, GB1 and GB2 for granule borders, IG, IG1
      and IG2 for intergranular lanes, LIG for a large intergranular
      lane, BP1 and BP2 for bright points, LMC1 and LMC2 for large
      magnetic concentrations, FT1 and FT2 for centers of thin and
      moderate fluxtubes.  FT3 is a strong fluxtube.  A cross in its
      center is called FT3-0 in the text; crosses to the left of
      center in the direction of the fluxtube periphery are called
      FT3-1, FT3-3, and FT3-5. The dotted line shows the selection of columns
      for the calculation of intensity and temperature
      fluctuations presented in Figure~\ref{Fig:dt_di}.
            }
         \label{Fig:images}
\end{figure}
%===========================================================================

%==================================================fig7=====================
\begin{figure}
   \centering
 \includegraphics[width=6cm]{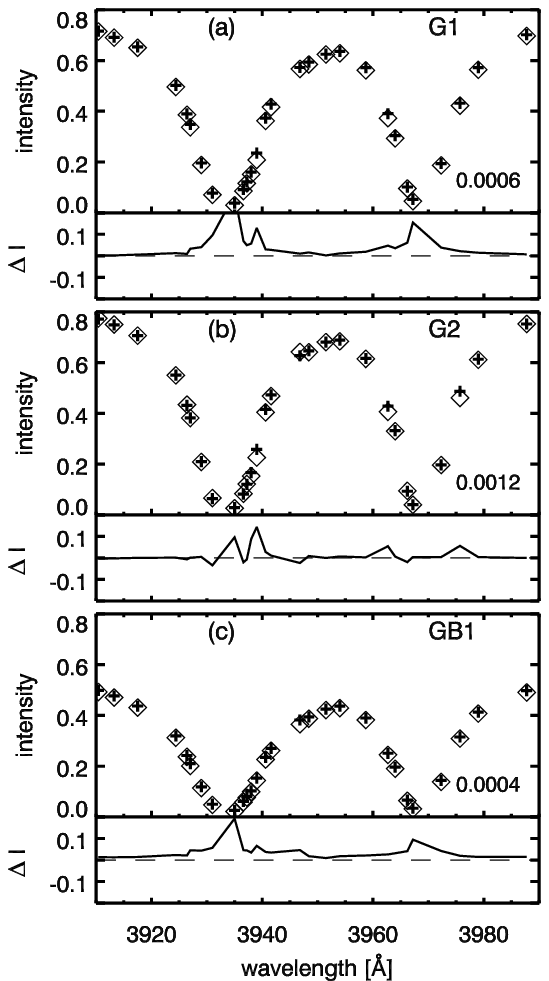}
 \includegraphics[width=6cm]{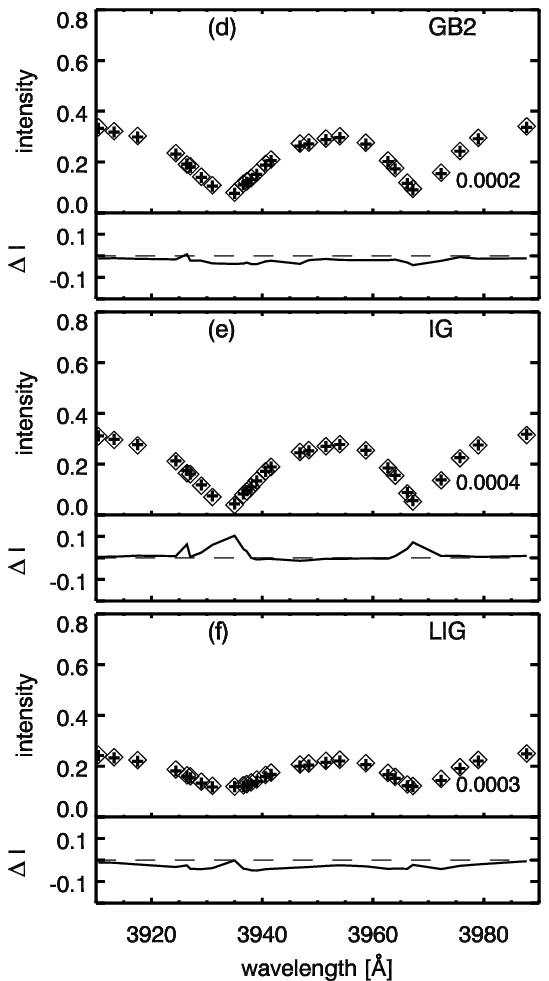}
 \caption[]{ Best-fit \HK\ profiles (crosses) compared to synthetic
   profiles from single 3DHD columns (rhombuses) for granule
   center (G1, G2), granule borders (GB1, GB2), an intergranular lane
   (IG), and a large intergranular lane (LIG).  The symbol
   and number coding are the same as in Figure~\ref{Fig:best_fit}.  }
         \label{Fig:p_fit_w}
\end{figure}
%===========================================================================

%======================================================fig8================
\begin{figure}
  \centering
     \includegraphics[width=8.cm]{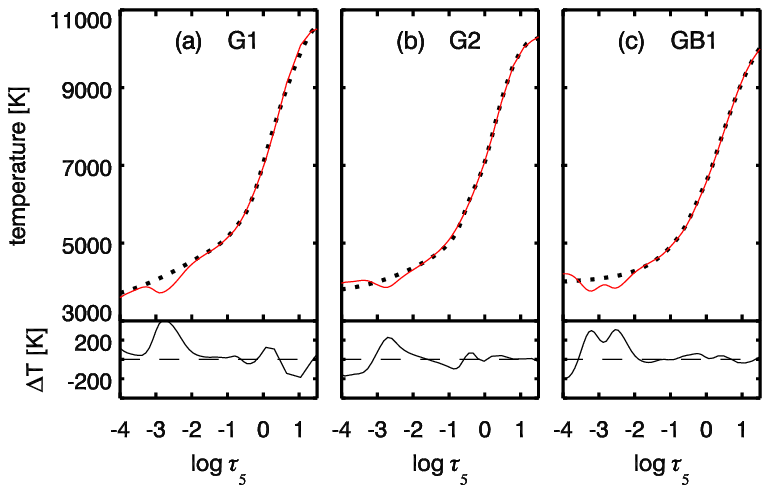}
     \includegraphics[width=8.cm]{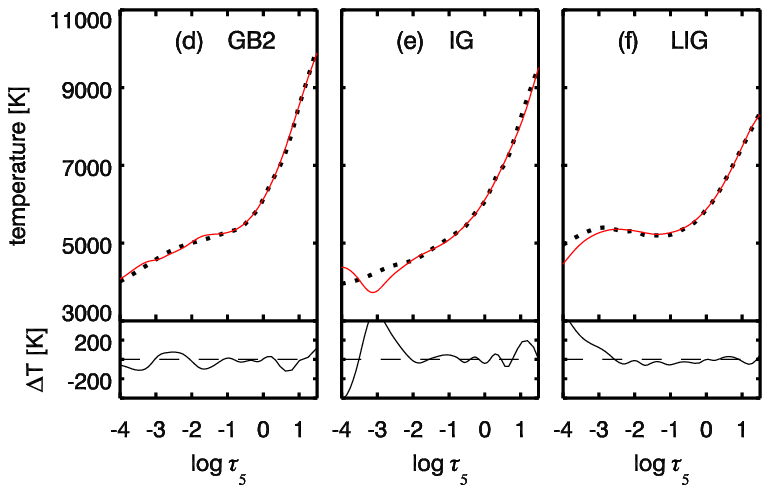}
     \caption[]{Best-fit temperature stratifications (dotted curves)
       derived from single 3DHD columns for the same locations as in
       Figure~\ref{Fig:p_fit_w}. Symbol and curve coding as in
       Figure~\ref{Fig:fit_mod}.  }
   \label{Fig:t_fit_w}
\end{figure}
%===========================================================================

%==================================================fig9=====================
\begin{figure}
   \centering
 \includegraphics[width=5cm]{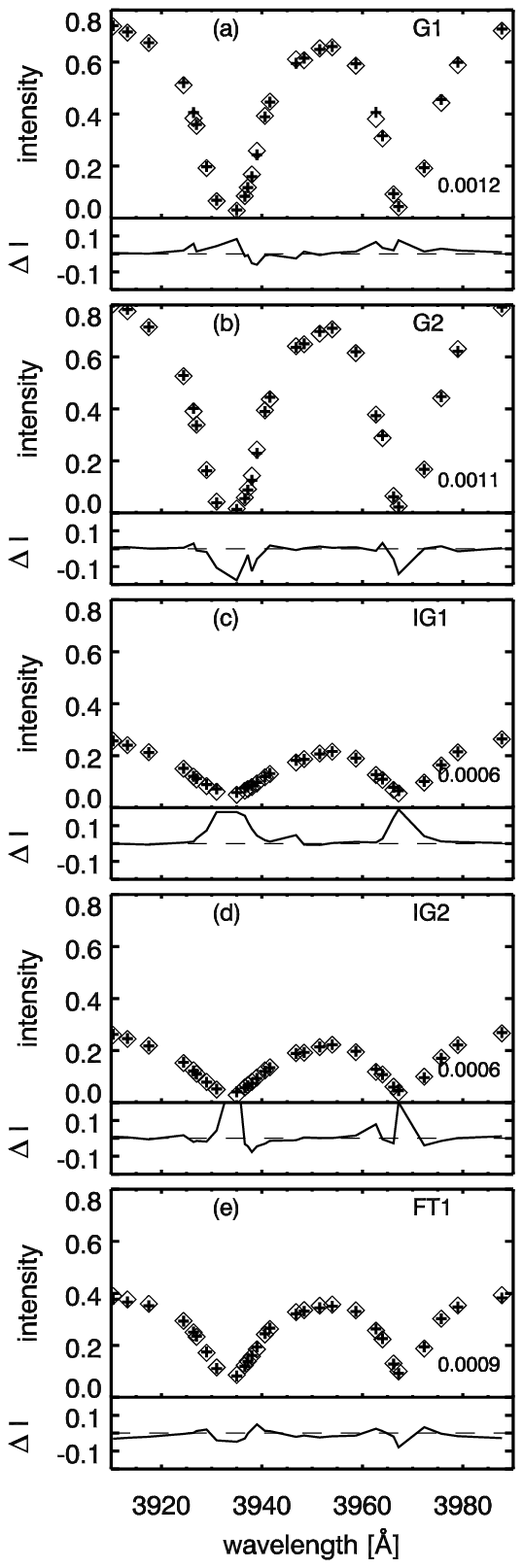}
 \includegraphics[width=5cm]{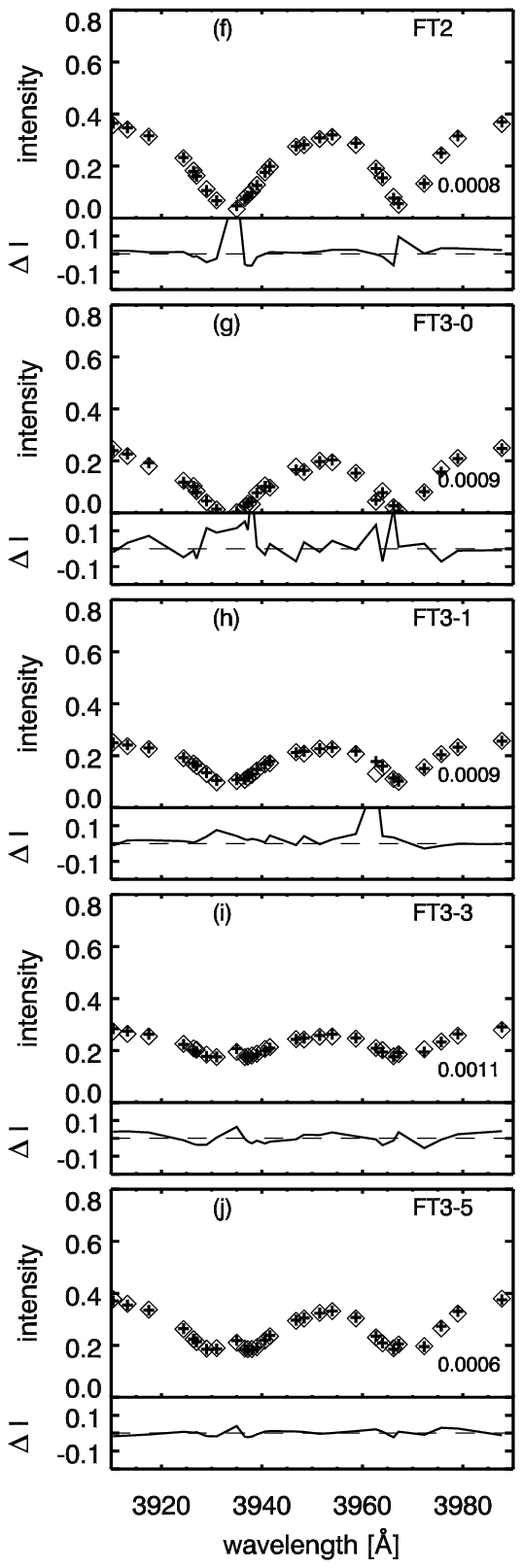}
 \caption[]{Best-fit \HK\ profiles from 2DMHD for granule center (G1,
   G2), intergranular lanes (IG1, IG2), thin fluxtube (FT1), moderate
   fluxtube (FT2), and four locations in the strong fluxtube (FT3-0,
   FT3-1, FT3-3, and FT3-5).  Symbol and curve coding as in
   Figure~\ref{Fig:best_fit}.  }
         \label{Fig:p_fit_g}
\end{figure}
%===========================================================================

%======================================================fig10=====================
\begin{figure}
  \centering
    \includegraphics[width=12.cm]{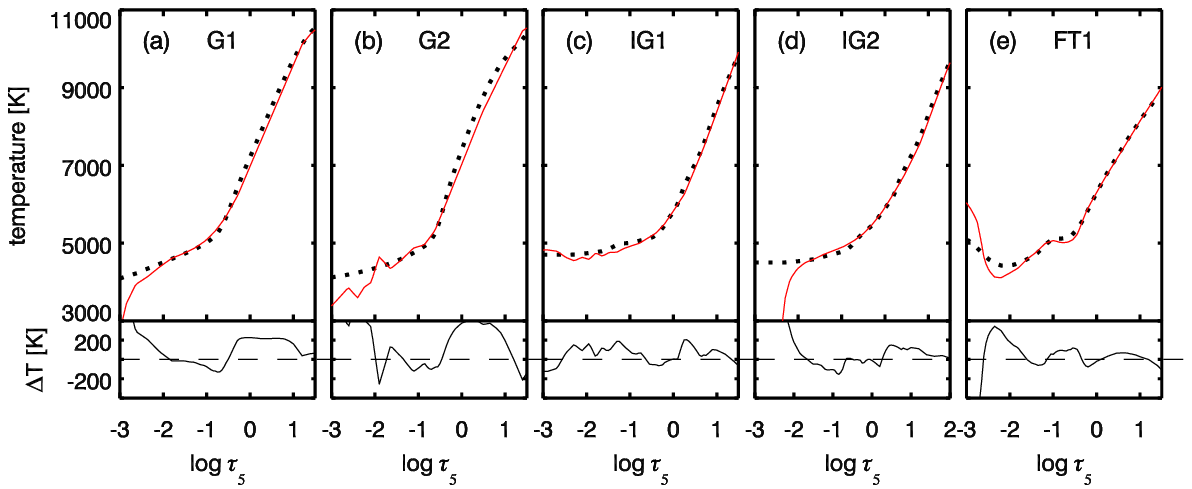}
    \includegraphics[width=12.cm]{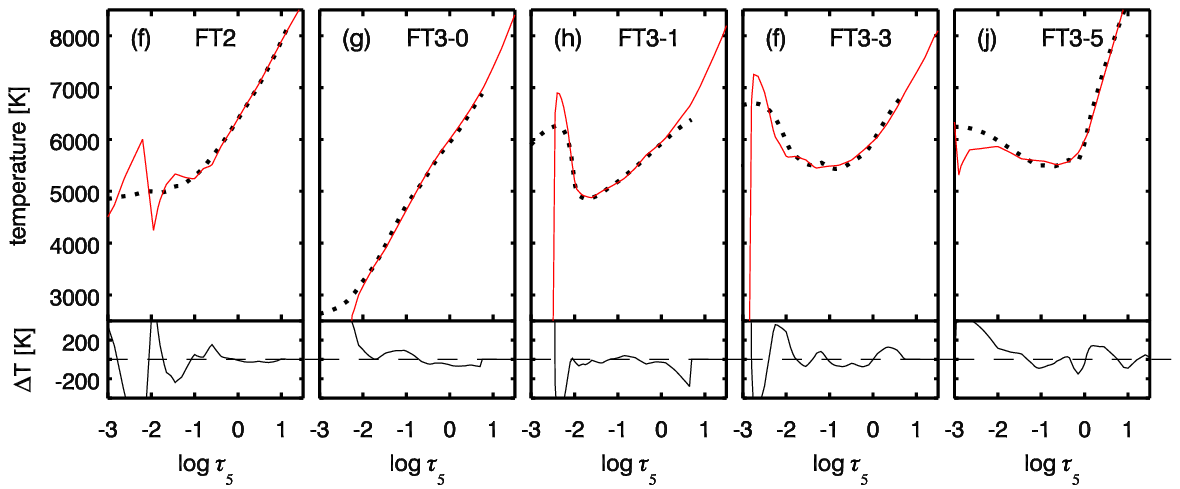}
    \caption[]{Best-fit temperature stratifications from 2DMHD.
      Symbol and curve coding as in Figure~\ref{Fig:fit_mod}.  }
   \label{Fig:t_fit_g}
\end{figure}
%===========================================================================

%==================================================fig11=====================
\begin{figure}
   \centering
   \includegraphics[width=5cm]{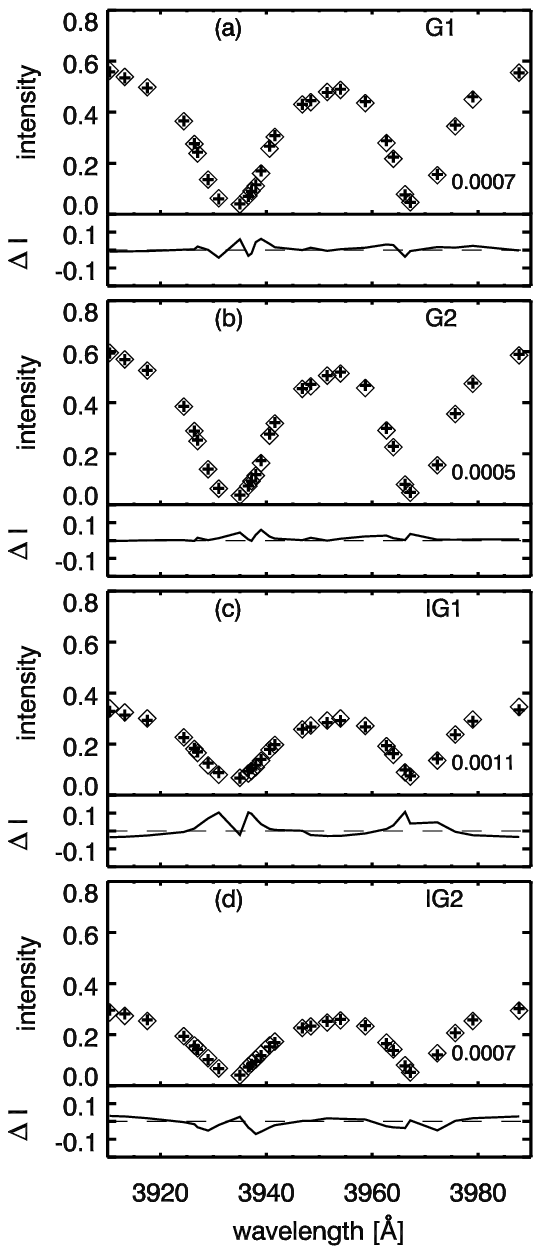}
   \includegraphics[width=5cm]{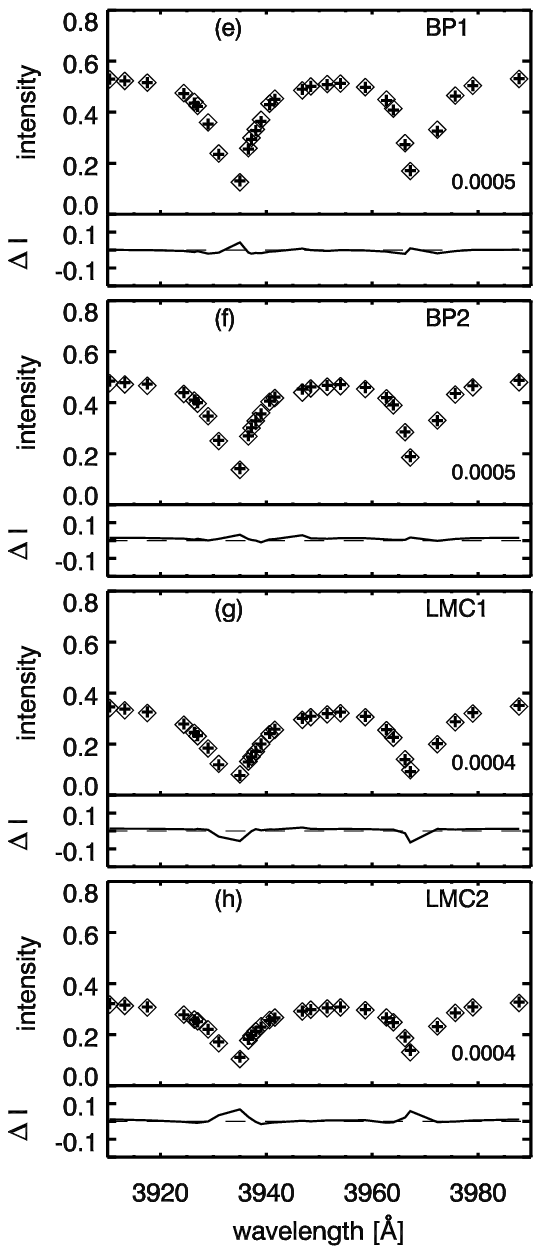}
   \caption[]{Best-fit \HK\ profiles from 3DMHD for granule center
     (G1, G2), intergranular lanes (IG1, IG2), bright points (BP1,
     BP2), and large magnetic concentrations (LMC1, LMC2).  Symbol and
     curve coding as in Figure~\ref{Fig:best_fit}. }
         \label{Fig:p_fit_s}
\end{figure}
%===========================================================================

%======================================================fig12================
\begin{figure}
  \centering

    \includegraphics[width=10.cm]{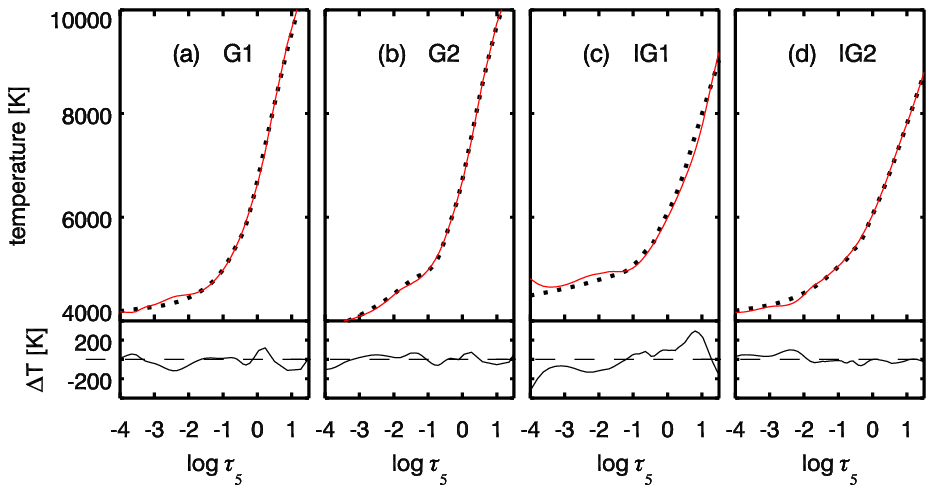}
    \includegraphics[width=10.cm]{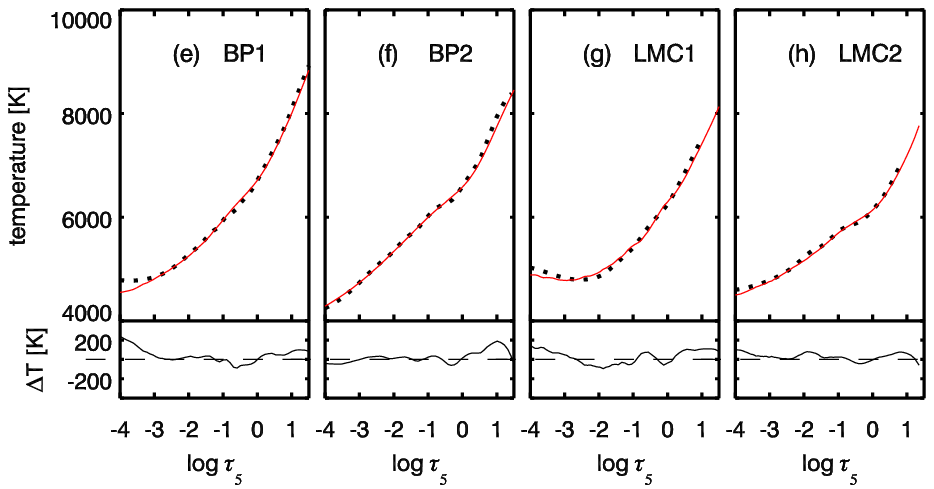}
    \caption[]{Best-fit temperature stratifications from 3DMHD.
      Symbol and curve coding as in Figure~\ref{Fig:fit_mod}.  }
   \label{Fig:t_fit_s}
\end{figure}
%==========================================================================

%%%%%%%%%%%%%%%%%%%%%%%%%%%%%%%%%%%%%%%%%%%%%%%%%%%%%%%%%%%%%%%%%%%%%%%%%%%%
\subsection{ Best-fit stratifications from spatially averaged synthetic
  profiles}
%%%%%%%%%%%%%%%%%%%%%%%%%%%%%%%%%%%%%%%%%%%%%%%%%%%%%%%%%%%%%%%%%%%%%%%%%%%%

This section presents best-fit results using the averaged \HK\
profiles. Figure~\ref{Fig:best_fit} shows the best-fit
synthesized \HK\ profiles with the spatially averaged \HK\
profiles from all  columns  of 3DHD, 2DMHD, and 3DMHD,
respectively. For each fitted profile the standard deviation is
\begin{equation}\label{eq_3}
  \sigma(I-I_{\rm original}) =\sqrt{\frac{1}{n}\sum_{i=1}^n (I(\lambda_i) -
  I_{\rm original}(\lambda_i))^2},
\end{equation}
where the wavelength index $i=1,...,n$ sample the wavelengths at
which spectrum is known, $n=28$, $I(\lambda)$ is the intensity
synthesized from the one-dimensional model, and $I_{\rm
original}(\lambda)$ is the average intensity over the
simulation. The largest values is $\sigma =0.0008$ for 2DMHD
which has the largest temperature spread and gradient. In
addition, the lower panels show residuals $\Delta I$ for quick
assessment of the fit quality. These are normalized to $I_{\rm
original}$ per wavelength as $\Delta I$~=~$(I-I_{\rm
original})/I_{\rm original}$.

Figure~\ref{Fig:fit_mod} compares the corresponding best-fit
temperature stratifications (dotted) to the simulation averages
(solid red).  The best agreement between the best-fit
stratifications and the simulation averages occurs over
$\log\tau_5 \is [0,-2]$. Higher up the best-fit temperatures lie
above the averages, especially for 2DMHD  due to the larger
thermal inhomogeneity of 2DMHD (Figure~\ref{Fig:t_av}b). Near
$\log\tau_5 \approx 0.5$ the best-fit stratifications lie above
the averages for all three simulations. Figure~\ref{Fig:fit_mod}
also shows that the temperature residual $\Delta T=T_{\rm
best-fit}- T_{\rm original}$ reaches a maximum of about 600~K
for the 2DMHD stratifications, probably due to the growth of
magnetic inhomogeneity.

%%%%%%%%%%%%%%%%%%%%%%%%%%%%%%%%%%%%%%%%%%%%%%%%%%%%%%%%%%%%%%%%%%%%%%%%%%%%
\subsection{ Best-fit temperature stratification from spatially
  resolved \HK\ profiles}
%%%%%%%%%%%%%%%%%%%%%%%%%%%%%%%%%%%%%%%%%%%%%%%%%%%%%%%%%%%%%%%%%%%%%
In this section a few columns corresponding to specific
fine-structure features (granule center, granule border,
intergranular lane, individual magnetic concentration) are
selected for best-fit modeling.

Figure~\ref{Fig:images} shows continuum images at
$\lambda=3910$~\AA\ constructed from the three simulations with
labels  at the selected locations. The small-scale magnetic
elements in 2DMHD were selected with the help of vertical cross
sections though the computational domain. In the 3DMHD
simulation, they were selected according to the data
of \inlinecite{2004ApJ...610L.137C}. % Carlsson++ G-band BP limg.

Figures~\ref{Fig:p_fit_w}, \ref{Fig:p_fit_g}, and
\ref{Fig:p_fit_s} show the best-fit results of the corresponding
resolved emergent profiles. The best-fit synthesized \HK\
profiles (crosses) and the profiles synthesized from the
individual columns of the numerical simulation (rhombuses) are
in close agreement.  The statistic estimations $\sigma(I-I_{\rm
original})$ in the lower-left corners and the residuals (in
relative units) in the lower panels show the fit quality.

Note the diversity between profiles from different types of
granular structure.  The steep wing slope becomes more gentle
from a granule center to its borders and to a lane
(Figure~\ref{Fig:p_fit_w}). The profile changes between the
centre and periphery of strong fluxtubes are seen in
Figures~\ref{Fig:p_fit_g}g--j.  The profiles from the bright
points (Figures~\ref{Fig:p_fit_s}e,f) are the narrowest.

Figures~\ref{Fig:t_fit_w}, \ref{Fig:t_fit_g}, \ref{Fig:t_fit_s}
show the best-fit temperature stratifications (dotted) and the
corresponding original 3DHD, 2DMHD, 3DMHD columns (red). In
general, the agreement is satisfactory.  Some differences occur
at temperature irregularities in the original stratifications
(e.g., Figures~\ref{Fig:t_fit_g}a,b,f,e). As is seen from these
figures, the temperature residuals $\Delta T$ are associated
with the irregularities except for the large $\Delta T$
(200--400~K) at $ \log \tau_5=0$ in Figure~\ref{Fig:t_fit_g}a,b.

%%%%%%%%%%%%%%%%%%%%%%%%%%%%%%%%%%%%%%%%%%%%%%%%%%%%%%%%%%%%%%%%%%%%%%%%%%%%
\section{Discussion}                                \label{sec:Discussion}
\begin{figure}
   \centering
   \includegraphics[width=8.cm]{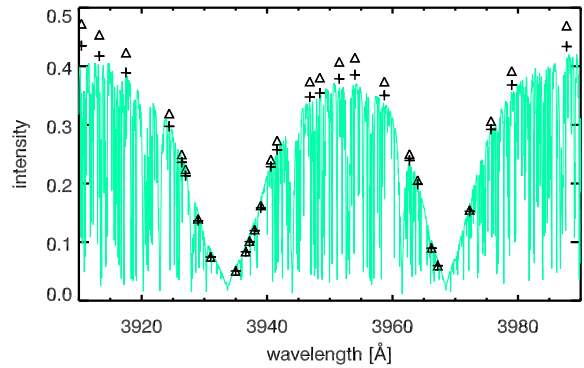}
   \caption[]{ Average \HK\ profiles  computed without (triangles)
     and  with (pluses) line-haze fudge factor
     of $c = 0.14$ in Eq.~(\ref{eq_1}) for the 3DHD simulation.
      The green observed profile  from the Brault-Neckel atlas.
}
         \label{Fig:int_fudg}
\end{figure}

%%%%%%%%%%%%%%%%%%%%%%%%%%%%%%%%%%%%%%%%%%%%%%%%%%%%%%%%%%%%%%%%%%%%%%%%%%%%
The discrepancies between the quiet-Sun atlas spectrum and the
averaged profile from the 3DHD simulation without magnetic
fields (Figure~\ref{Fig:hk_fit}a) may  have two causes. The
first is the neglect of the ``line haze'' which affects the
atlas profile but is not accounted for in the simulated
profiles. Figure~\ref{Fig:int_fudg} shows that inclusion of a
``line haze'' in the \HK\ profile synthesis can significantly
decrease the intensity in the \HK\ wings. Unfortunately,
choosing an appropriate height-dependent fudge factor is not
trivial.  The second reason is that the temperature fluctuations
in the simulation may not reproduce those in the real Sun.
Evaluation of the simulation correctness is beyond this paper.

The differences between the averaged profiles derived from the
three simulations (Figure~\ref{Fig:hk_fit}a) can be due to the
difference in the temperature inhomogeneities, gradients, and
magnetic field density (${<}|B|{>}$).  Figure~\ref{Fig:t_av}b
shows the largest temperature spread for 2DMHD which represents
magnetogranulation with ${<}|B|{>} = 400$~G in strong network.
The large temperature spread in the higher layers of the 2DMHD
increases the inner-wing intensities. It is caused by the
presence of many magnetic network elements.
Figure~\ref{Fig:t_av}b shows many sharp temperature drops and
rises caused by the strongest magnetic fields within the
fluxtubes and at their hot walls at the peripheries.  This
follows from the analysis of vertical cross sections through the
computational domain in the 2DMHD sequence.  An example is
Figure~10 of Paper~1. The temperature gradient of 2DMHD is also
the steepest one in Figure~\ref{Fig:hk_fit}b.

The temperature fluctuations observed as granulation of the
solar surface are created by penetrating convective flows. The
thermodynamic properties of the granulation are described by
\inlinecite{1998ApJ...499..914S} in detail.  The temperature and
density of the plasma near the surface have approximately
bimodal distributions. The intergranular lanes have low
temperature, very low ionization, and high density. The granules
have high temperature, high ionization, and low density.  The
granules occupy larger area than the lanes. Below the surface
they occupy $\approx \frac{2}{3}$ of the total area.  Here the
radiative cooling proceeds very rapidly. Since the $H^-$ opacity
is dominant and very temperature sensitive  $(\approx T^{10})$,
the steepest vertical temperature gradient results near the
surface in the granules.

%==============================================================fig13=======

%==========================================================================

%==========================================================fig14========
\begin{figure}
   \centering
   \includegraphics[width=7.cm]{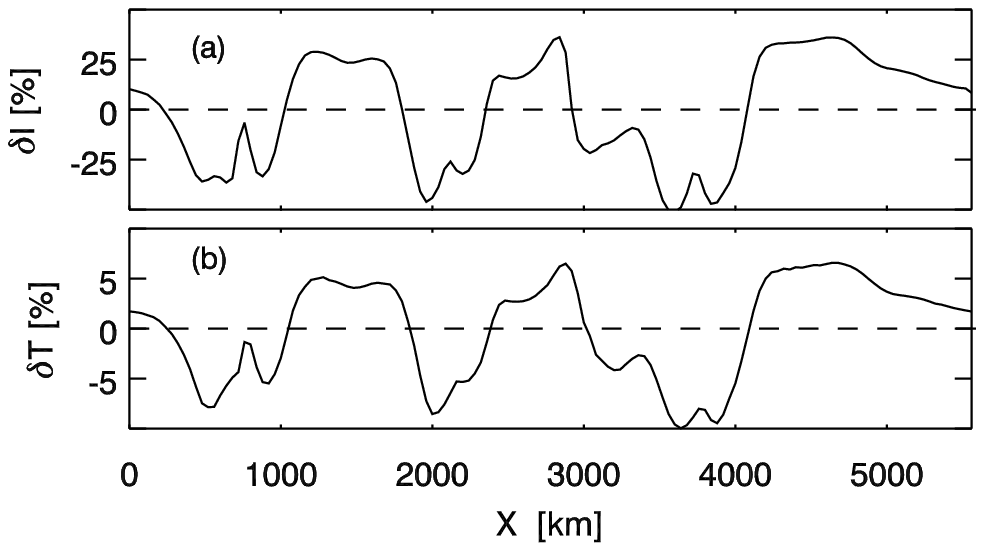}
   \caption[]{ Fluctuations of the emergent
     intensity ($\delta I$) in the far
     wing of the  \CaIIH\ line at $\lambda=3910$~\AA\,.  The
     temperature fluctuations ($\delta T$)  at the optical depth equal to unity
     inferred from a set of columns are signaled by the dotted
     line in a 3DHD  snapshot  (Figure~\ref{Fig:images}a).
              }
         \label{Fig:dt_di}
\end{figure}
%===========================================================================

The presence of magnetic field in a simulation also affects the
granulation. \inlinecite{2000KFNT...16...99G} showed that the
granulation pattern differs for magnetogranulation: the granules
are larger and the shearing instability is stronger in the
absence of magnetic fields. The field has a stabilizing effect
-- the granules are smaller and there are no pronounced
horizontal shears.  The distribution of granules in
magnetogranulation differs from the nonmagnetic granulation by a
greater contribution of bright features, which are associated
with bright points in flux tubes.  Therefore the granular
contribution to the emergent intensity can also differ between
field-free granulation and magnetogranulation depending on the
magnetic field density.

In spatial averaging the temperature fluctuations have different
effect on the temperature average and on the average emergent
intensity, due to the nonlinear dependence of the Planck
function on temperature (\opencite{2011ApJ...736...69U}).
Indeed, Figure~\ref{Fig:dt_di} shows intensity and temperature
fluctuations ($ \delta I=(I-{<}I{>})/{<}I{>}$ and $ \delta T=
(T- {<}T{>})/{<}T{>}$) computed for a set columns through of a
3DHD snapshot (marked by a dotted line  in the
Figure~\ref{Fig:images}a). Here  ${<}T{>}$ is the mean
temperature  along the selected cut in this snapshot  at optical
depth equal to unity. ${<}I{>}$ is the emergent intensity in the
far wing of the \CaIIH\ line at $\lambda=3910$~\AA\ averaged
along the selected cut.  The intensity fluctuations  differ
significantly in percentage from the temperature fluctuations.
The relatively small temperature fluctuations $\delta T = 5$\%
produce large intensity fluctuations $\delta I = 50$\%.  The
intensity contributions to the average profile from the columns
located within the granules are larger than from the columns
located in intergranular lanes. This means that the averaged
surface intensity from a simulation, i.e., averaged over the
columns of a multi-dimensional simulation, is larger than the
average intensity calculated from the averaged simulation.

%==========================================================fig15========
\begin{figure}
   \centering
   \includegraphics[width=6.cm]{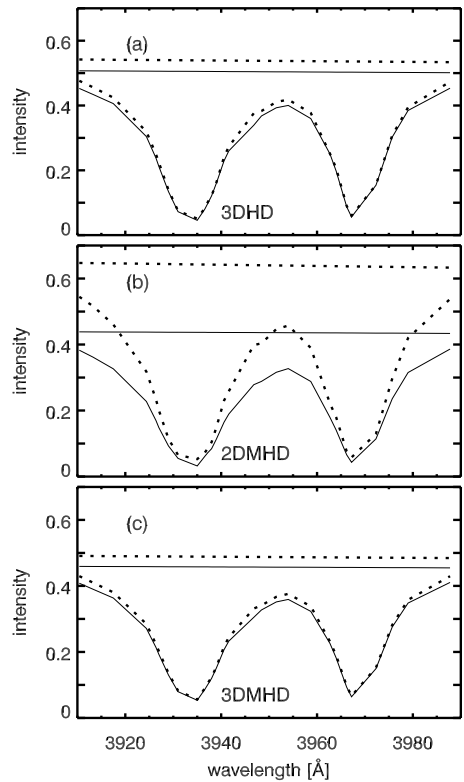}
   \caption[]{Averaged line profiles synthesized from the 3DHD, 2DMHD,
     3DMHD  simulations (dotted  curves) and the profiles synthesized from their
     average temperature stratifications (solid), as well as
     the corresponding continua.
   }
         \label{Fig:cont}
\end{figure}
%===============\============================================================

%==========================================================fig16========
\begin{figure}
   \centering
   \includegraphics[width=12.cm]{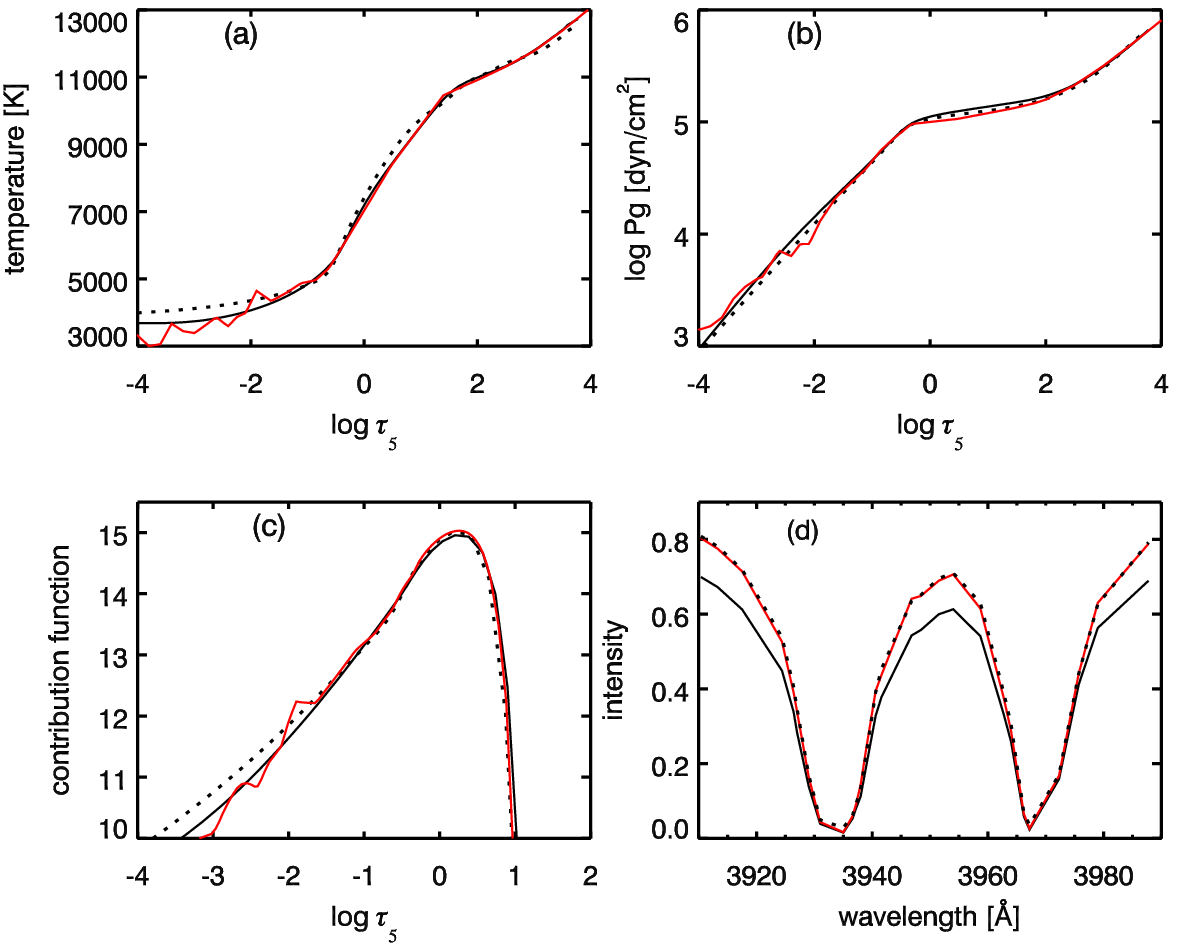}
   \caption[]{Temperature (a) and gas pressure  (b) stratifications for a
     granule (Figure~\ref{Fig:t_fit_g}b) as well as contribution
     functions (c) to the emergent intensity at $\lambda=3910$\,\AA \ and
     line profiles (d):    original (red curves),  smoothed (black),
     and best fit (dotted). Contribution functions are given
     in logarithmic units.  }
         \label{Fig:t_gran}
\end{figure}
%===============\====================

Figure~\ref{Fig:cont} shows the excess intensity between such
averaged \HK\ profiles and the profiles synthesized from each
spatially-averaged  simulation. The continuum excess in relative
units at $\lambda=3910$~\AA\ amounts to 7\% for 3DHD, 38\% for
2DMHD, and 7\% for 3DMHD. The intensity excess in the far wings
is close to the continuum excess. Figure~\ref{Fig:fit_mod}
demonstrates that the best-fit procedure leads to overestimation
of the temperature at continuum formation heights.  The maximum
temperature excesses are 194~K (3DHD), 619~K (2DMHD), and 48~K
(3DMHD).  The larger the temperature inhomogeneities, the higher
is the best-fit continuum and the larger the temperature excess
of the best-fit average temperature stratification. Thus, in the
deepest layers of the quiet photosphere the one-dimensional
modeling of the mean atmosphere give overestimation of about
200~K, whereas in the strong network  regions the overestimation
can reach 600~K.

The best-fit stratifications for the discrete fine-structure
elements derived from the synthetic \HK\ spectra in
Figures~\ref{Fig:t_fit_w}, \ref{Fig:t_fit_g}, \ref{Fig:t_fit_s}
recover the original temperature stratifications satisfactorily.
Notice that the fitting procedure applied here recovers
successfully both a temperature stratification which decreases
monotonically with height and stratifications with a mild or
steep temperature rise in higher layers
(Figures~\ref{Fig:t_fit_w}f, \ref{Fig:t_fit_g}e,h,f,j,
\ref{Fig:t_fit_w}d, \ref{Fig:t_fit_g}c,
\ref{Fig:t_fit_s}c,e,f,g,h). The question arises whether it is
possible to identify a temperature rise from the \HK\ profiles
alone.  Figures~\ref{Fig:p_fit_w}, \ref{Fig:p_fit_g},
\ref{Fig:p_fit_s} show that \HK\ profiles calculated for
temperature-rise stratifications  have higher intensity near the
line core the for a stratification with decreasing temperature.
Probably, this diagnostic can be employed indicator of an
atmospheric temperature rise.

In Figures~\ref{Fig:t_fit_w}, \ref{Fig:t_fit_g},
\ref{Fig:t_fit_s} the temperature residuals  $\Delta T=T_{\rm
best-fit}-T_{\rm original}$ reflect the intrinsic uncertainties
of the method used. For example, Figure~\ref{Fig:t_fit_g}b shows
large (about 400~K)  deviations  from the original models  
at $\log\tau_5=0$, i.e., large deficiencies of the method. Tests
showed that adding radiation pressure, magnetic pressure, and
turbulent pressure in Eq.~(\ref{eq_2}) makes no significant
difference. Since the chemical composition is the same for the
profile synthesis and the best-fit modeling it does not affect
the results. The real cause for such deviations may be a effect of thermal
irregularities in the vertical direction on the best-fit
stratification. The effect is tested in Figure~\ref{Fig:t_gran}.
The smooth temperature stratification (black curve) differs from
the original (red curve) at locations where there are sudden
temperature changes with depth. Such thermal irregularities are
not recovered by the method used. In Figure~\ref{Fig:t_gran}b
the gas pressure $P_g(h)$ (black) derived from the smooth $T(h)$
with Eq.~(\ref{eq_2}) has excesses compared to the original one
(red) at the locations corresponding to sharp temperature
gradients. The synthesized \HK\ profile (black) with the smooth
$T(h)$ therefore does not match the original profile (red) in
Figure~\ref{Fig:t_gran}d. The corresponding black contribution
function  for the intensity at $\lambda=3910$~\AA\
(Figure~\ref{Fig:t_gran}c) is slightly lower  than the red one
due to increase of the damping ($\gamma_{\rm wdw}\approx P_g$)
and the line opacity $\kappa(T, P_g)$. It causes lower emergent
intensity in the far wings. To obtain a best line-profile fit
the temperature increases and the pressure decreases (see dotted
curves). Therefore the best-fit temperature (dotted curve)
demonstrates the excess about 400~K in Figure~\ref{Fig:t_gran}a.

Complex non-monotonic temperature stratification such as in
Figure~\ref{Fig:t_fit_g}d,e,f also cannot be reproduced by the
manual fitting procedure used here. In addition, in the manual
fitting  the initial $T(h)$ and especially $P_g(h)$
stratifications in the deepest layers must be as close as
possible to the real ones.  So, for example, in the case of
magnetic elements it is not possible to obtain best-fit profiles
with the HSRA-SP-M model as initial guess, whereas the PLANEW
fluxtube model gives good results. These are shortcomings of the
manual technique which diminish when using automatic inversion
codes
(e.g., \opencite{1997ApJ...478L..45B}; % bellot
\opencite{2000A&A...358.1109F}) % frutiger
that search for a best fit by means of a Marquardt nonlinear
least-squares algorithm.  They can locate the minimum of the
merit function more precisely than a manual procedure. The
retrieved temperature stratification by the automatic inversion
codes is largely independent of the choice of the initial guess
atmosphere.  Note, it is a limitation for the manual technique,
but not for the \HK\ wings photospheric diagnostics.  Therefore,
it is desirable to develop such inversion codes for fitting \HK\
profiles.

%%%%%%%%%%%%%%%%%%%%%%%%%%%%%%%%%%%%%%%%%%%%%%%%%%%%%%%%%%%%%%%%%%%%%%%%%%%%
\section{Conclusion}                                \label{sec:conclusion}
%%%%%%%%%%%%%%%%%%%%%%%%%%%%%%%%%%%%%%%%%%%%%%%%%%%%%%%%%%%%%%%%%%%%%%%%%%%%
The extended wings of \CaII\ \HK\ supply relatively easy
diagnostics to extract quantities that describe vertical
stratification of the solar photosphere. I have calibrated such
one-dimensional modeling using three numerical simulations of
solar fine structure, containing granular convection with
different amounts of magnetic flux.  I obtained best-fit
temperature stratifications from synthetic \HK\ wing profiles
computed from these simulations. Quantitative evaluation permits
the following conclusions.

The different effect of thermal inhomogeneities on the average
temperature stratification and on the average emergent intensity
due to the non-linear Planck function sensitivity leads to
overestimation of temperatures derived in such one-dimensional
modeling. For quiet-Sun conditions this overestimation reaches
about 200~K in the continuum formation layers. The larger the
inhomogeneity, the larger the excess. The largest overestimation
occurs for one-dimensional modeling at field strengths similar
to the 2DMHD simulation, and can be as large as 600~K for strong
magnetic fields  when average profiles are analyzed.  Such
temperature overestimation may also occur in the upper
photosphere due to large temperature contrasts between hot
fluxtube walls and cool fluxtube insides.

The \HK\ wing diagnostics provide better recovery of the actual
temperature stratifications within discrete fine structure elements
such as a granule center, an intergranular lane, a bright point, a
fluxtube, a large magnetic concentration.  Here errors occur only
where there are sharp vertical gradients and non-smooth irregularities
in the original temperature stratification.

The range of effective formation of the emergent intensity in
the wings of  \CaII\ \HK\  extends from $\log \tau_5 =-2$ to
0.2. Beyond this depth range the reliability of the  \HK\ wing
diagnostics is less.

In conclusion, one-dimensional \HK\ wing modeling using
high-resolution data is a reliable technique to construct
best-fit models for small-scale features in the solar photosphere.

%%%%%%%%%%%%%%%%%%%%%%%%%%%%%%%%%%%%%%%%%%%%%%%%%%%%%%%%%% ACKNOWLEDGEMENTS
\begin{acknowledgements}
  I am much indebted to Rob Rutten for suggesting this analysis and
  for extensive improvement of the text, to Sven Wedemeyer-B\"ohm, Robert
  Stein, and {\AA}ke Nordlund for the use of the 3DHD and 3DMHD
  simulations, and to Jorrit Leenaarts, Mats Carlsson and Rob Rutten
  for providing these.
  I am grateful to  an anonymous referee for the many helpful
   comments and important suggestions to improve significantly
    the presentation of this paper.
\end{acknowledgements}

%%%%%%%%%%%%%%%%%%%%%%%%%%%%%%%%%%%%%%%%%%%%%%%%%%%%%%%%%%%%%%%% REFERENCES
\bibliographystyle{spr-mp-sola}
\bibliography{reference}
%%%%%%%%%%%%%%%%%%%%%%%%%%%%%%%%%%%%%%%%%%%%%%%%%%%%%%%%%%%%%%%%%%%%%%% END

\end{article}
\end{document}